\documentclass[aps,twocolumn,floats,prd,nofootinbib,superscriptaddress,showpacs,10pt]{revtex4-1}

\usepackage{amssymb}
\usepackage{amsmath}
\usepackage{verbatim}
\usepackage{mathrsfs}
\usepackage{amsfonts}
\usepackage{latexsym}

\begin{document}

\title{Scalar field \textit{vs} hydrodynamic models  in the 	 homogeneous isotropic cosmology}

\author{V.~I.~Zhdanov}\email{valeryzhdanov@gmail.com}
\affiliation{Astronomical Observatory, Taras Shevchenko National University of Kyiv} %

\author{~S.~S.~Dylda}\email{tunerzinc@gmail.com}
\affiliation{Physical Faculty, Taras Shevchenko National University of Kyiv} %

\begin{abstract}
We study relations between hydrodynamical (H) and scalar field (SF) models of the dark energy in the homogeneous isotropic Universe. The focus is on  SF described by the Lagrangian with the canonical kinetic term within   spatially flat cosmology. We analyze  requirements that  guarantee  the same cosmological history for the SF and H models at least for special solutions.  The differential equation for the SF potential is obtained that ensures such  equivalence of the SF and H-models. However, if the ``equivalent'' SF potential is found for given  equation of state (EOS) of the H-model, this does not mean that all solutions of this SF-model have corresponding  H-model  analogs. In this view we derived a condition that guarantees an ``approximate equivalence'', when there is a  small difference between  energy-momentum tensors of the models. The ``equivalent'' SF potentials and corresponding SF solutions for  linear  EOS are found in an explicit form; we also present examples with more complicated EOSs.
\end{abstract}


\maketitle

\section{Introduction} \label{Introduction}
Observations show \cite{observations}  that about 70\% of the average  mass density in the Universe owes to the dark energy (DE) which drives the  acceleration of the cosmological expansion. It is widely assumed that some form of DE or its constituents that dominated in the very early Universe must have a dynamical nature owing to an action of unknown physical fields  and/or due to  modifications of the General Relativity  \cite{DE_review, gravity_tests}. 
Theories with scalar fields (SF) occupy an important sector of this area (see, e.g., \cite{DE_review, gravity_tests, Constraints_inflation,bamba_et_al_2012,nojiri-odintsov-2011}). Though observational data restrict some of the SF models \cite{Constraints_inflation}, there is still a considerable uncertainty  in their choice,  not to mention the  revision of the underlying gravitational theory  \cite{gravity_tests,bamba_et_al_2012}. The  abundance of various cosmological models draws attention to unifying schemes and interrelations between competing dark energy candidates \cite{bamba_et_al_2012, nojiri-odintsov-2011, Oikonomou_etal_2017} that can be used to explain observational data.

To this end, the phenomenological  hydrodynamic approach is often used  \cite{bamba_et_al_2012, nojiri-odintsov-2011, eos_models, hydrodynamic_perturbations,p=p_of_e_phi}. It is well known that the matter in the spatially homogeneous and isotropic  Universe can be described by means of the energy-momentum tensor of an ideal fluid. Under certain conditions SF models allow for a hydrodynamic (H) description with some ``equivalent'' equation of state (EOS)  beyond homogeneity as well \cite{hydrodynamic_perturbations}. The H analogs of the SF models typically involve such phenomenological parameters as the EOS parameter,  effective sound speed,  adiabatic sound speed, which can be limited in view of available astronomical data \cite{hydrodynamic_perturbations, hydrodynam_parameters_recent}. The transition from simplest hydrodynamical EOSs to  SFs and vice versa deals with rather unusual models, whereas in the spirit of Occam's razor, it would be desirable to restrict the choice of the SF Lagrangian to  canonical one, which is more familiar from the point of view of particle physics. This is possible within the approach of papers \cite{bamba_et_al_2012, nojiri-odintsov-2011, eos_models, zhdanov},   which treat the equivalence problems by direct comparing solutions for the cosmological scale factor and the energy density in the homogeneous  isotropic Universe. As distinct from these papers, we propose a differential equation for the SF potential in closed form guaranteeing some equivalence of H and SF models.  We note, however, that in any approach, the H-SF correspondence is not universal;  this is well known though not always clearly stated.  A typical situation is that two different models mimic each other for some area of the original data, but they have different solutions outside this area.

We study the H-SF equivalence on the basis of equality of the corresponding energy-momentum tensors  (Section \ref{general}).
This problem becomes more  complicated if we impose some additional conditions either on the form of the EOS, or on the SF Lagrangian.  We focus on the relationship between  the H-model and the SF-model with the  canonical kinetic term and a self-interaction potential for the real SF. We call this ``restricted\footnote{because we deal with the special form (\ref{canonical_l}) of Lagrangian} equivalence", in contrast to the case, when no such  restrictions are imposed. As a result, the equivalence considerations deal with some restrictions on the initial data (Section \ref{statement}).

Then we consider a case, when the relations that guarantee some kind of equivalence of SF and H models  are valid approximately. In this case the equations of H and SF models can lead to different energy-momentum tensors (and correspondingly different evolution equations), and the question is when this difference remains small, provided that it is small at the initial moment.
We derived conditions for such approximate equivalence (Section \ref{comparison}).

The results are applied to the linear EOS, including situation near the phantom line (subsection \ref{linear}), to some non-linear EOS known from papers \cite{bamba_et_al_2012, zhdanov} (subsection  \ref{nonlinear}), and to a simple two-parametric EOS (subsection \ref{two-parametric_eos}). Here we present examples showing when one can speak about an equivalence between the H and SF models.

\section{General considerations}
\label{general}
General Lagrangian $L(X,\varphi)$ for the real SF $\varphi$ with $X=\frac{1}{2}\varphi,_{\mu}\varphi^{,\mu}$ and space-time metric $g_{\mu\nu}$ yields the  energy-momentum tensor\footnote{We use the signature $(+,-,-,-).$}
\begin{equation}
	\label{sf-e-m-tensor}
	T_{\mu \nu }^{(sf)} = \frac{2}{\sqrt { - g} }\frac{\partial}{\partial g^{\mu \nu }} \left[ \sqrt { - g} L \right] = \frac{\partial L }{\partial X}\, \varphi _{,\mu } \varphi
	_{,\nu } - g_{\mu \nu } L \, ,
\end{equation}
that can be equated to the  energy-momentum tensor of an ideal fluid
\begin{equation}
\label{h-e-m-tensor}
T_{\mu \nu }^{(h)} = hu_\mu u_\nu -p g_{\mu \nu } ,
\end{equation}
where  $h = e + p$  is the specific enthalpy, $p$ is the pressure, $e$ is the invariant energy density and $u^{\mu}$ is the four-velocity of the fluid. It is assumed that some EOS is known that relates the pressure to the other parameters of the problem: $p=P(e,\varphi)$.

We have $T_{\mu \nu }^{(sf)} =T_{\mu \nu }^{(h)}$  if
\begin{equation}
\label{h1} p=L(X,\varphi),\quad
h = 2X\partial L/\partial X,\quad X>0.
\end{equation}
At the points where $X$ changes its sign (i.e., $\varphi_{,\mu}$ is not  timelike), the hydrodynamical interpretation is no longer valid.  The relations (\ref{h1}) yield an ordinary differential equation with respect to $L(X,\varphi)$, where $\varphi$  is involved as a parameter:
 \begin{equation}
 \label{h2}
 E(L,\varphi) = 2X\partial L/\partial X-L.
 \end{equation}
The solution $L$ of  (\ref{h2})  exists in case of rather a general EOS;  this solution  contains an arbitrary function of $\varphi$. Additional constraints  that ensure equality of (\ref{sf-e-m-tensor}) and (\ref{h-e-m-tensor}) for all values of $\varphi$ and its derivatives are outlined in Appendix \ref{general_case}. These constraints are fulfilled identically in case of an homogeneous isotropic Universe to be discussed further.

However, under additional restrictions on the functions $L(X,\varphi)$ and/or $E(p,\varphi)$ in (\ref{h2}) the solution of this equation for all $X,\varphi$  may not exist. For example, if we want to define the EOS parametrically from  (\ref{h1}), then a general Lagrangian $L$ cannot yield the  barotropic EOS, because in this case the   right-hand sides of (\ref{h1}) may depend on two independent  variables $X$ and $\varphi$.

In this view we shall require that equations (\ref{h1},\ref{h2}) be satisfied not for arbitrary hydrodynamical and/or SF variables, but only for certain cosmological solutions in the isotropic homogeneous Universe. We wonder, is it possible to compare H-model with the SF-model, if the SF Lagrangian has the canonical form
  \begin{equation}
  \label{canonical_l}
 L = X - V(\varphi ) .
\end{equation}
 Then equations (\ref{h1},\ref{h2}) yield
 \begin{equation}
 \label{equiv_1}
 e=X+V(\varphi),\quad p(e,\varphi)=X-V(\varphi).
 \end{equation}
If we demanded that (\ref{equiv_1})  be fulfilled for all variables $(e, \varphi)$ or $(X, \varphi)$, we would have very special EOS $p=e-2V(\varphi)$, whereas in case of  another equations of state the relations  (\ref{equiv_1}) cannot be identities. However, we deal with the unique Universe, so in fact we do not need that the ``equivalence'' conditions be satisfied for all possible values of the variables that enter EOS and/or  Lagrangian. If we compare different cosmological models, then the main question is when they predict the same observational data, when they mimic each other etc.  

We say that there is an equivalence of H and SF models, if both predict the same Hubble diagram and, consequently, the same Hubble parameter $H(z)$ as a function of  the redshift $z$. In this case we have the same $H(t)$ as a function of the cosmological time $t$, yielding the same\footnote{Up to unessential constant factor in  case of the spatially flat cosmology.}  dependence of the cosmological scale factor $a(t)$.

Within the homogeneous isotropic cosmology, a solution $a(t),\varphi(t)$ of the SF-model\footnote{This is specified in the next section.} yields $e(t),p(t)$ as a parametric representation of EOS due to (\ref{equiv_1}) and vice versa; this approach is often considered (see, e.g. \cite{bamba_et_al_2012, nojiri-odintsov-2011}). The main difference of the present paper is that we are looking for a direct criterion on the SF potential $V(\varphi)$, which yields the same cosmological history as in the case of H-model with the  prescribed EOS.

\section{Statement of the problem}
\label{statement} 
We consider the spatially flat cosmology described by the Friedmann-Lemaitre-Robertson-Walker metric
\begin{equation}
\label{eq1}
ds^2 = g_{\mu \nu } dx^\mu dx^\nu = dt^2 - a^2\left( t \right)\left[ {d\chi
	^2 + \chi^2 dO^2} \right]\,.
\end{equation}
\noindent
It should be noted that the supposition of spatial flatness agrees  with observations \cite{observations} and is perfectly explained in the framework of widely known ideas of inflation in the early Universe \cite{Constraints_inflation}. In case of the Universe filled with an ideal fluid we have the Friedmann equations (spatially flat case)
\begin{equation}
\label{eq2}
\frac{d^2a}{dt^2} = - \frac{4\pi }{3}a\left( {e + 3p} \right),
\end{equation}
\begin{equation}
\label{eq3}
H^2 = \frac{8\pi }{3}e ,
\quad
H = a^{ - 1}da / dt 
\end{equation}
($G = c = 1)$ .
One more (hydrodynamical) equation
\begin{equation}\label{fluid}
 \dot{e}+3H (e+p)=0
\end{equation}
 also follows from (\ref{eq2},\ref{eq3}); on the other hand, (\ref{eq2}) follows from (\ref{eq3},\ref{fluid}).
 Further we use (\ref{eq3},\ref{fluid}) as the independent equations taking in mind that they must be supplemented by an equation of state.

In case of the isotropic homogeneous Universe filled with uniform scalar field $\varphi=\varphi(t)$; $X=\dot{\varphi}^2/2$  the evolution equations corresponding to (\ref{canonical_l}) are 
\begin{equation}
\label{eq2SF}
\frac{d^2a}{dt^2} = - \frac{8\pi }{3}a\left( {\dot\varphi^2- V(\varphi)} \right),
\end{equation}
\begin{equation}
\label{eq3SF}
H^2 = \frac{8\pi }{3}e_f ,
\end{equation}
where $e_f\equiv \dot\varphi^2/2  + V(\varphi)$ is the field energy density, and the field equation is 
\begin{equation}
\label{eq8SF}
\ddot {\varphi } + 3H\dot {\varphi } + V'(\varphi ) = 0 .
\end{equation}
Analogously, these equations are not independent and we use further (\ref{eq3SF}, \ref{eq8SF}) as the evolution equations of the SF-model with the initial conditions
\begin{equation}
\label{varphi_initial}
\dot \varphi(t_0)=\dot \varphi_0,\quad   \varphi(t_0)=  \varphi_0.
\end{equation}

If (\ref{eq3},\ref{fluid}) and (\ref{eq3SF},\ref{eq8SF}) are fulfilled with the same $H(t)$, then 
\begin{equation}
\label{constraint_a}
e=e_f\equiv \dot\varphi^2/2  + V(\varphi),
\end{equation}
and substituting this into (\ref{fluid}) and using (\ref{eq8SF}), we get 
\begin{equation}
\label{constraint_b}
\dot{\varphi}^2=e+p.
\end{equation}
Analogously, on account of (\ref{constraint_a}, \ref{constraint_b}) equations (\ref{eq3SF},\ref{eq8SF})  yield (\ref{eq3},\ref{fluid}). Conversely, (\ref{eq3},\ref{fluid}) and  (\ref{eq8SF},\ref{constraint_a})  yield (\ref{constraint_b}). The relations (\ref{constraint_a},\ref{constraint_b})  are necessary for the equivalence of H and SF models. 
These conclusions do not depend on either $p=P(e)$ or $p=P(e,\varphi)$.  However, the statement of the  initial value problem of H-model and its comparison with SF-models looks somewhat different in case of (i) one-parametric and (ii) two-parametric EOS.

(i) {\it Barotropic EOS: $p=P(e)$}. 
The H-model is defined by equations (\ref{eq3},\ref{fluid}) with 
the initial condition 
\begin{equation}
\label{e=e_0}
e(t_0)=e_0 
\end{equation}

(ii) {\it Two-parametric EOS}. We shall see below that the requirement of equivalence of H and SF models imposes severe limitations on cosmological solutions. In order to generalize the discussion and verify that the limitations are not due to the one-parametric form (i), we  consider  EOS that contains two parameters. Following \cite{p=p_of_e_phi} we suppose that $p=P(e,\phi)$;  this generalization can be used to construct phenomenological models of the dark energy. Obviously, this demands that dynamical equation  for the additional variable $\phi$ must be also involved in the H-model. Now we wonder, is it possible to describe the solutions of this model with the help of some scalar field $\varphi$ alone, without using the hydrodynamic variables? The very first step in this direction and the most economic way within our "restricted" approach is to suppose that $\phi=\varphi$ obeys the same  equation (\ref{eq8SF}). Therefore, we assume that the equations of the H-model include  (\ref{eq3},\ref{fluid},\ref{eq8SF}) with  corresponding  initial conditions (\ref{varphi_initial},\ref{e=e_0}). 
Our formal aim is to find criteria for existence of the H-model solution $e(t)$ and the SF-model solution $\varphi(t)$  with the same $H(t)$,  such that $e_f(t)\equiv e(t)$. 

Obviously,  considering  (ii)  of the H-model with two-parametric EOS and its comparison to the SF-model differs from considering (i), in particular, because we have different dimensions of the space of initial data. However, the mathematics we deal with below is formally the same and the equivalence criterion (\ref{eq16}) derived below is applicable both to (i) and (ii). So  further we work with (ii), having in mind the reservation concerning the difference of (i) and (ii). 

We assume  $h(e,\varphi)$ to be a continuously differentiable function of $e,\varphi$. Further for brevity we denote
\[G(x,y)\equiv x^2-h(x^2/2+V(y),y).\]
Using this function,  in view of the relations (\ref{constraint_a},\ref{constraint_b}), we have 
\begin{equation}
	\label{constraint}
	G(\dot{\varphi}(t),\varphi(t))=0.
\end{equation}
As we have seen,  this condition along with (\ref{constraint_a}) ensures that both H-model and SF-model lead to the same Hubble diagram (at least for specially chosen initial data). In this sense we speak about ``restricted\footnote{Because we deal with the restricted Lagrangian (\ref{canonical_l}).} equivalence'' of  H and SF-models. 
The condition (\ref{constraint}) must be fulfilled  for initial data (\ref{e=e_0}) as well:
\begin{equation}
\label{constraint_data}
	G(\dot{\varphi}(t_0),\varphi(t_0))=0.
\end{equation}

Also, we shall consider deviations from  equation (\ref{constraint}); in this case we consider the function 
\begin{equation}
\label{g_of_t}
g(t) = G(\dot {\varphi(t) },\varphi(t)) .
\end{equation}

It should be noted that for fixed $V(\varphi)$,  $h(e,\varphi)$ it is generally impossible to satisfy (\ref{constraint_data}) with  $\forall \dot \varphi_0,  \varphi_0$; this relation  singles out a particular solution to equation (\ref{eq8SF}). Therefore most of solutions of the SF-model cannot be H-model solutions.

After the comments about the initial data we proceed to conditions for the potential, which must be fulfilled  to ensure (\ref{constraint}). We suppose that the function $h(e, \varphi)$ is known.   The problem we are interested in can be formulated as follows.

A. Let  $\varphi(t)$ be a solution of (\ref{eq3SF},\ref{eq8SF}). What are sufficient conditions for $V(\varphi )$ so as to ensure $g(t)\equiv 0$ at least for special initial data (\ref{varphi_initial}) satisfying (\ref{constraint_data})? 

After finding potential $V(\varphi)$ that solves the problem (A) for special initial data satisfying (\ref{constraint_data}), it is natural to ask about   another solutions of the SF-model with the same potential, which do not satisfy (\ref{constraint_data}). 

B. Let for some $V(\varphi)$  there are solutions $\varphi(t), \bar \varphi(t)$ of  equations (\ref{eq3SF},\ref{eq8SF}), and
\begin{equation}
\label{problem_B}
G(\dot{\bar \varphi}(t),\bar \varphi(t))=0,\quad
G(\dot{\varphi} (t),\varphi (t))\ne 0.
\end{equation}
 What we can say about $g(t)$ in (\ref{g_of_t})? 
 
 If $ \varphi(t_0)$ satisfies (\ref{constraint_data})  approximately, will this approximation work for  $t > t_0 $? If yes, we can say that we have an ``approximate equivalence'' of H and SF-models.

\section{Comparison of S and H models}
\label{comparison}
The equation (\ref{g_of_t}) can be solved with respect to  $\dot \varphi^2$. With this aim we introduce function $\Theta (V,g,\varphi)$, which is defined as a solution of the equation
\begin{equation}
\label{eq12}
\Theta = g +  h\left( {\Theta/2 + V, \varphi} \right)  .
\end{equation}
Uniqueness of solution of (\ref{eq12}) can be easily established if, for $\varepsilon=const > 0$ (arbitrarily small), 
\begin{equation}
\label{uniqueness}
\frac{\partial h}{\partial e}  \le 2 - \varepsilon.
\end{equation}
The uniqueness follows from consideration of  $\zeta(\vartheta)=\vartheta - h\left( {\vartheta/2 + V, \varphi} \right)$, which is monotonically increasing function of $\vartheta$. Then $\zeta(\vartheta)$ takes the value  $\zeta(\vartheta)=g$  only once, therefore we have a unique  solution $\vartheta=\Theta(V,g,\varphi)$ of (\ref{eq12}). A sufficient condition of existence is  $h(V,\varphi)\ge -g$, because in this case  $\zeta(0)\le g$ and $\zeta(\vartheta)\to \infty$ as $\vartheta\to \infty$ due to (\ref{uniqueness}); so in virtue of continuity of $\zeta(\vartheta)$ there exists the solution $\vartheta$ of (\ref{eq12}). In case of $g=0$ this sufficient condition is simply the requirement for the  positive   specific enthalpy. 
Note that (\ref{uniqueness}) means $\partial P/\partial e\le 1-\varepsilon<1$, which avoids superluminal speed of sound.

We also introduce $E(V,g,\varphi) =  \Theta (V,g,\varphi) / 2 +
V$ that satisfies the equation
\begin{equation}
\label{eq13}
E - \frac{1}{2}h\left(E,\varphi\right) = \frac{1}{2}g + V .
\end{equation}
Further we consider solutions of (\ref{eq13}) such that $e=E(V,g,\varphi) > 0,\quad \dot \varphi^2=\Theta (V,g,\varphi) > 0$.

After differentiation of (\ref{g_of_t}) and in view of (\ref{eq3SF},\ref{eq8SF}) we get
\begin{equation}
\label{eq14a}
\dot {g} = - \dot {\varphi} \left[ \dot {\varphi }\sqrt {24\pi e}   \left(2 - \frac{\partial h}{\partial e}\right)   +\frac{\partial h}{\partial \varphi}+
2\frac{dV}{d\varphi}  \right],
\end{equation}
where $h\equiv h(e,\varphi)$ and we denote $e = \frac{1}{2}\dot {\varphi }^2 + V(\varphi )>0$.

If we require $ {g} \equiv 0$, then, for $\dot {\varphi}\ne 0$, we have 
\begin{equation}
\label{eq16}
\frac{dV}{d\varphi }+\frac{1}{2}\frac{\partial h}{\partial \varphi} + S\left( {2 - \frac{\partial h}{\partial e}} \right)\sqrt {6\pi E_0 \Theta_0} =0 ,
\end{equation}
where $h=h(e,\varphi)$, $e=E_0 (V,\varphi) \equiv E(V,0, \varphi)$, $\Theta_0 (V,\varphi) \equiv \Theta (V,0,\varphi)$, $S=\mbox{sign}(\dot {\varphi })$, and we used (\ref{eq12},\ref{eq13}).
Note that equation (\ref{eq16}) is a formal consequence of (\ref{eq14a}) only for those $\varphi$ that belong to the range of solutions $\varphi(t)$ of (\ref{eq8SF}). 

The differential equation (\ref{eq16}) for the potential $V(\varphi)$ is a sufficient condition to have (\ref{constraint}) provided that   $g(t_0)=0$.
Thus, the problem (A) of equivalence is reduced to the equation for potential $V(\varphi)$ in closed form, which, however, is different for different signs of $\dot {\varphi }$.

In case of a barotropic EOS $h=h(e)$ equation (\ref{eq16}) is simplified to the form
\begin{equation}
\label{eq16_barotropic}
 \frac{dV}{d\varphi }=- S\left( {2 - \frac{\partial h}{\partial e}} \right)\sqrt {6\pi E_0 (V,\varphi)\Theta _0 (V,\varphi)}  .
\end{equation}
In virtue of (\ref{eq13}) we have for $V=E_0-h(E_0)/2$. Substitution to  (\ref{eq16}) yields  more compact equation
\begin{equation}
\label{eq25}
\frac{dE_0 }{d\varphi } = - S\sqrt {24\pi 	\,E_0 \,h(E_0 )} .
\end{equation}  
For given EOS, equations (\ref{eq16_barotropic},\ref{eq25}) allow to find $V(\varphi)$ such that certain modes of cosmological evolution $H(t), e(t)=e_f(t)$ can be obtained by means of either H-model or SF-model. This, however, does not apply to all possible solutions to this SF-model, in particular, when $\dot \varphi$ changes its sign.
 From (\ref{eq16_barotropic}) it follows that the potential $V(\varphi)$   must be a  monotonically increasing function provided that we consider an interval where $\dot {\varphi }<0$. This includes, e.g., the case of the slow-roll modes of the chaotic inflation.

\section{Initial data not satisfying (\ref{constraint_data})}
Now we proceed to (B). Let $V=V(\varphi)$ satisfies (\ref{eq16}) and $\bar\varphi(t), \varphi(t)$ satisfy (\ref{problem_B}). We shall consider  an interval of $t$, where $S=\mbox{sign}[\dot {\varphi }(t)]=\mbox{sign}[\dot {\bar \varphi}(t) ]$ is constant.

Using  (\ref{g_of_t},\ref{eq12},\ref{eq13}), we substitute expressions for $e,\dot \varphi$ into equation (\ref{eq14a}) to have a first order ordinary differential equation with respect to $g(t)$:
\begin{equation}
\dot {g} = - \dot \varphi\left\{ {S\,K(\varphi,g) \left[ {2 - \frac{\partial h}{\partial e}} \right] + \frac{\partial h}{\partial \varphi}+ 2\frac{dV}{d\varphi}} \right\},
\end{equation}
where in the r.h.s.  $K(\varphi,g)= \sqrt {48\pi e (e-V)}$,  $h=h(e,\varphi)$, $e=E(V,g,\varphi)$, $V=V(\varphi)$.

Denoting
\begin{equation}
\label{eq18}
D(\varphi,g) = S\,K(g,\varphi) \left[ 2 - \frac{\partial h}{\partial e} \right] + \frac{\partial h}{\partial \varphi} \, ,
\end{equation}
where $h=h(e,\varphi)$, $e=E(V(\varphi),g,\varphi)$, in virtue of  (\ref{eq16}), which is true for any $\varphi$,   we have
\[
2\frac{dV}{d\varphi}=-D(\varphi,0)
\]
yielding
\begin{equation}
\label{g_dot}
 \dot g =- g\,   \dot {\varphi }  R(\varphi,g)
\end{equation}
where $ R(\varphi,g) = g^{-1}  \left\{D(\varphi,g)-D(\varphi,0)\right\}$ is a regular function.
From (\ref{g_dot}) we get 
\[
g(t) = g(t_0)\exp \left\{ { -  \int\limits_{t_0}^t {ds} \,  {\dot{\varphi }(s)}    R\left[ {\varphi (s),g(s)} \right]} \right\} . \]
The behavior  of $g(t)$ depends on the monotonicity sign of $D(\varphi,g)$ as a function of $g$, which defines the sign of $R(\varphi,g)$. If
\begin{equation}
\label{R_monotony}
S\,R(\varphi,g)  > 0,
\end{equation}
then  $|g(t)|\le |g(0)|$ for $t>0$ and we arrive at the approximate equivalence for a sufficiently small initial $g(0)$.
Moreover, if $ \dot
	{\varphi}(s)   R(\varphi,g)\ge \beta> 0, \, \beta=\mbox{const}$, then we have $g(t)\to 0$ for $t\to \infty$  exponentially.

One  can estimate the sign of (\ref{R_monotony})
under supposition of differentiability of (\ref{eq18}).
 Equations (\ref{eq12},\ref{eq13}) yield
\begin{equation}
\label{deq13dVde}
\frac{\partial E}{\partial g}=\frac{\partial \Theta}{\partial g} =  \left[ {2 - \frac{\partial h}{\partial e}} \right]^{-1}.
\end{equation}
The monotonicity condition (\ref{R_monotony})  transforms to
\[
S\,\frac{\partial D}{\partial g}=  \frac{\sqrt{24\pi}}{\sqrt {E \,\Theta }} \left\{\frac{\Theta}{2} +E-\frac{E\Theta}{2-\partial h/\partial e} \frac{\partial^2 h}{\partial e^2} \right\} +
\]
\begin{equation} \label{eq20}
+\frac{S}{2 - \partial h/\partial e} \frac{\partial^2 h}{\partial e\partial \varphi}
>0,
\end{equation}
where  $h=h(e,\varphi)$,  $e=E(V(\varphi),g,\varphi)$,  $\Theta= \Theta(V(  \varphi),g, \varphi)$.
Since $R(\varphi,0)=\partial D/\partial g$ for $g=0$,  if this inequality is fulfilled for $g=0$, then we have the ``approximate equivalence'', i.e. at least for sufficiently small $g(0)$ we have  $|g(t)|\le |g(0)|$ for $t>0$ and in this sense we say that $\varphi(t)$ well approximates $\bar\varphi(t)$ on interval where the signs of $\dot {\bar{\varphi}}(t)$ and $\dot \varphi(t)$ are equal.

\section{Examples}
\label{examples}
\subsection{Linear equation of state}
\label{linear}
Now we shall consider an example with a concrete equation of state. The simplest one is the linear barotropic EOS:
\begin{equation}
\label{eq21}
h(e) = \xi (e - e_0 ) + h_0 = \xi e - \eta ,\quad \eta = \xi e_0 - h_0 .
\end{equation}

Solutions of equations (\ref{eq12}, \ref{eq13}) 
are
\[
\Theta _0(V) = \frac{2(\xi V - \eta )}{2 - \xi },\quad E_0 (V) = \frac{2V - \eta }{2 - \xi };
\]
they are uniquely defined for $\xi \ne 2$, so we assume this condition instead of (\ref{uniqueness}). Equation (\ref{eq16_barotropic})  takes on the form
\begin{equation}
\label{eq22}
\frac{dV}{d\varphi } = - S_1 S\sqrt {24\pi
\,\xi \,\left[ {\left( {V - \frac{2 + \xi }{4\xi }\eta } \right)^2 -
\left( {\frac{2 - \xi }{4\xi }\eta } \right)^2} \right]} ,
\end{equation}
$S_1= \mbox{sign}(2-\xi) $. 

For $\xi > 0$ the solution of (\ref{eq22}) that obeys inequalities $\Theta_0\ge 0,E_0\ge 0$ is
\begin{equation} \label{potential_1}
V(\varphi ) = \frac{(2+\xi)\eta}{4\xi }   +  \frac{(2 - \xi)}{4}  \frac{|\eta| }{\xi }  \cosh \left[
{2\alpha (\varphi - {\psi _0} )} \right]  ,
\end{equation}
$\alpha = \sqrt {6\pi \xi }$,   under condition  that
\begin{equation} \label{additional_1}
\mbox{sign}[\dot \varphi\,(\varphi -{\psi}_0 )] = - 1,
\end{equation}
${\psi}_0$ is an integration constant. The other options that do not yield positive $\Theta_0$ and $E_0$ have been discarded. For $0<\xi<2$ the potential (\ref{potential_1}) has minimum at $\varphi =\psi_0$; for $\xi>2$ the potential  is unbounded from below.  

The particular solution $\bar\varphi(t)$ of the SF-problem (\ref{eq3SF},\ref{eq8SF}) with the initial data satisfying (\ref{constraint_data}) can be found from the first order differential equation (\ref{constraint}); it generates the solution of the H-problem  $e(t)=\bar e_f(t) \equiv\dot{\bar\varphi}^2/2 +V(\bar{\varphi})$.

Consider, e.g., the case of $\eta>0$; here  (\ref{constraint}) on account of (\ref{additional_1}) leads to the equation
 \[
 \dot{\bar\varphi}=-\sqrt{\eta} \sinh \left[
 {\alpha (\bar \varphi - {\psi _0} )} \right],
 \]
yielding two solutions 
\begin{equation}
\label{particular_solution}
   \bar \varphi(t)=\psi_0 \pm \frac{1}{\alpha}\mbox{arsinh}\left\{ \sinh[\alpha\sqrt{\eta}(t-t_1)] \right\}^{-1},
\end{equation}
$ t>t_1=const$. 

Correspondingly 
\[
e(t)=\bar e_f(t)=\frac{\eta}{\xi}\left\{\coth[\sqrt{6\pi\xi\eta}
(t-t_1)]\right\}^2
\]
is the solution of (\ref{eq3},\ref{fluid}), $\xi>0,\eta>0$. 

For any $S_1$ (\ref{particular_solution}) represents the monotonically decreasing/increasing function  that never reaches  $\varphi=\psi_0$ and $\bar e_f(t) $  never reaches the  value $e=\eta/\xi$.  The other solutions  of  (\ref{eq3SF},\ref{eq8SF}) with the same $V(\varphi)$ but not satisfying (\ref{constraint_data}) at $t=t_0$,  do not fulfill (\ref{eq3},\ref{fluid}) with the same $h(e)$ (\ref{eq21}) with   $e(t)=e_f(t)$.  For example, for $\xi<2$ the solutions of (\ref{eq8SF}) that oscillate near the minimum of the potential cannot be described by the H-model (\ref{eq21}):  this would contradict to (\ref{additional_1}) after passing either the turning point $\dot \varphi=0$ or the point $\varphi=\psi_0$. 

There is some freedom in the choice of the solution of (\ref{eq22}), which can be used, if we study a correspondence not between models with fixed $h(e)$ and/or $V(\varphi)$, but between families of potentials and equations of state. Suppose that for initial data (\ref{varphi_initial})  we have  $\dot\varphi_0^2\ne\eta   \sinh^2\left[ \alpha (\varphi_0-\psi_0)  \right] $, i.e. (\ref{constraint_data}) is not valid. However, by transforming parameters $\xi,\eta$ of the EOS (\ref{eq21}) or  $\psi_0$ of the potential, one can find some new values of these parameters  to satisfy  (\ref{constraint_data}) and to find the other special solution of SF-model that corresponds to H-model.

The condition (\ref{eq20}) is fulfilled at least for closest to (\ref{particular_solution}) solutions because $\Theta_0/2+E_0>0$. Therefore  $\bar \varphi(t)$ well approximates such solutions on whole intervals where   $\mbox{sign}[\dot {\varphi }(t)]=\mbox{sign}[\dot {\bar \varphi}(t) ]$. Though in case of (\ref{potential_1}) is is easy to study the qualitative behavior of solutions of  (\ref{eq3SF},\ref{eq8SF}); it is easy to see that $\varphi(t) \to \psi_0$ and, in view of continuous dependence of solution on ant finite interval upon the initial data, the deviation of $\varphi(t)$ from $\bar \varphi(t)$ will be small for all $t>t_0$, provided that it is small at $t=t_0$.
 
Now we proceed to the case $\xi < 0$, $\eta<0$. This example\footnote{The case  $\xi < 0$, $\eta>0$ is ruled out in view of the requirement $\Theta_0>0, E_0>0$.}  is unlikely to be of cosmological significance, but it illustrates  problems that can arise when in the course of evolution there are points with zero energy density.  From  (\ref{eq22}) we obtain the periodic potential
\begin{equation} \label{h_0_2}
 V(\varphi )= \frac{\eta (2 + \xi) }{4\xi }  + \frac{\eta(2 - \xi)}{4\xi} \cos \Phi ,
\end{equation}
where $\Phi=2\alpha (\varphi -{ \psi }_0 )$, $\alpha = \sqrt {6\pi |\xi| }$; the additional condition for (\ref{eq22})  to be fulfilled is $\dot {\varphi }\,\sin\Phi>0$. On account of this condition and restricting ourselves to the range $\Phi\in (0,\pi)$, using (\ref{constraint})  we have the   solution $\bar \varphi(t)$ of (\ref{eq8SF}): 
\begin{equation}\label{termination}
\bar\varphi(t)=\psi_0+\frac{1}{\alpha}\arccos\left\{\tanh[\alpha\sqrt{|\eta|}(t_1-t)]\right\} ,
\end{equation}
for $t<t_1=const$. This relation describes  the SF evolution from $\bar\varphi=\psi_0$ to 
$\bar\varphi = {\psi}_0+\pi/(2\alpha)$. There is no analytic continuation\footnote{There is a loss of Lipschitz continuity in (\ref{eq8SF}).} of this solution for $t>t_1$. Correspondingly,  
\begin{equation}\label{termination2}
e(t)=\bar e_f(t)=\left|\frac{\eta}{\xi}\right|\left\{\tanh[\sqrt{6\pi|\xi\eta| }(t-t_1)] \right\}^2
\end{equation}
is the solution of hydrodynamical equations (\ref{eq3},\ref{fluid}) for $t<t_1$, where $h(e)$ is given by (\ref{eq21}). There is a trivial extension of (\ref{termination2}) for  $t>t_1$.

At last, consider the important case $\xi<0, \eta = 0$, yielding the famous ``Big Rip'' hydrodynamical solution  \cite{caldwell_doomsday}. In this case RHS of (\ref{eq22}) is not real, there is no non-trivial solution for the potential and there is no canonical SF counterpart.

\subsection{Example of a nonlinear EOS}
\label{nonlinear}
Consider the barotropic EOS \cite{bamba_et_al_2012, zhdanov}
\begin{equation}\label{nonlin_baro}
h(e) = \xi e\left[ {1 - \left( {e / e_0 } \right)^\mu } \right] ,
\end{equation}
where $\mu>0$. We are looking for possible solutions of (\ref{eq25}).

For $0<\xi <2$ the conditions for existence, uniqueness and positivity of $\Theta_0, E_0$ of (\ref{eq16}) can be verified directly using the solution below. Equation (\ref{eq25}) yields 
the solution $E_0 (\varphi ) = e_0 \left( {\cosh \Phi } \right)^{ - 2 / \mu }\le e_0$,  
where $\Phi = \mu \sqrt {6\pi  \xi   } \left(
{\varphi - {\psi}_0 } \right)$, $\psi_0$ is an integration constant, under condition that $\mbox{sign}\left[ {\dot {\varphi }\,(\varphi - \psi _0 )} \right] =1$.  Then we have the potential (cf. \cite{zhdanov})
\[
V(\varphi ) = E_0 - \frac{1}{2}h(E_0 )=\]
\begin{equation}
\label{V_nonlin}
= \frac{e_0 }{2\left( {\cosh \Phi }
	\right)^{2(1 + 1 / \mu )}}\left[ {\xi + \left( {2 - \xi } \right)\cosh
	^2\Phi } \right].
\end{equation}
The SF-model with this potential has the particular solution $\bar \varphi(t)$ satisfying (\ref{constraint});  correspondingly $e_f(t)$ satisfies the equations (\ref{eq3}, \ref{fluid}) of the H-model. Equation (\ref{constraint}) takes on the form  
 \[
\dot {\bar\varphi}=\frac{\sqrt{\xi e_0} \sinh \bar\Phi}{(\cosh\bar\Phi)^{1+1/\mu}}\,,\,\,  \bar\Phi = \mu \sqrt {6\pi  \xi   } \left(
{\bar\varphi - {\psi}_0 } \right).
  \]
For $\varphi >\psi_0$ this means that  $\bar \varphi(t)$  descents  down the potential hill to the right of $\psi _0$; it grows logarithmically and $\bar e_f(t)\to 0$ for $t\to \infty$. For $t\to -\infty$ we have $\bar \varphi(t)\to \psi_0$ and $\bar e_f(t)\to e_0$.

The condition (\ref{eq20}) for $g=0$ yields
\[
\frac{\Theta_0}{2}+E_0+\frac{\xi \mu (\mu+1)\Theta_0}{2-\xi+\xi(\mu+1)(E_0/e_0)^\mu} \left(\frac{E_0}{e_0}\right)^{\mu}>0 .
\]
This is always fulfilled for $0<\xi <2$ thus guaranteeing that 
$ \varphi(t)\approx \bar \varphi(t)$ on appropriate intervals in case of small deviation of the initial data. 

Note that the SF-model with $\xi>0$  does not admit   divergent solutions  like the ``Big Rip'' \cite{caldwell_doomsday} of the  hydrodynamical counterpart.

The case $\xi < 0$ is possible for $e > e_0 $; equation (\ref{eq25}) yields   
$E_0 (\varphi ) = e_0 \left( {\cos \Phi } \right)^{- 2 / \mu}\ge e_0$ for $\left|\Phi \right| < \pi / 2$,  $\Phi = \mu \sqrt {6\pi  |\xi|   } \left(
{\varphi - {\psi}_0 } \right)$. We have a potential pit with infinite walls (with a periodic continuation):
\begin{equation}\label{V_sin}
V(\varphi ) = \frac{e_0 }{2\left( {\cos \Phi } \right)^{2(1 + 1 / \mu
		)}}\left[ {\xi + \left( {2 - \xi } \right)\cos ^2\Phi } \right],
\end{equation}
and the SF solutions with this potential lead to the same evolution of $H(t)$  as the hydrodynamical ones if
\[
\mbox{sign}\left[ {\,\dot {\varphi }\,(\varphi-\psi_0)} \right] = - 1.
\]
In this view, equation  (\ref{constraint}) for the corresponding particular solution of (\ref{eq8SF})  yields  
\[
\dot {\bar\varphi}=-\frac{\sqrt{|\xi| e_0 } \sin \bar\Phi}{(\cos\bar\Phi)^{1+1/\mu}}\,.
\]
The scalar field slides off the wall and tends to the potential minimum with energy $e=e_0$, $\varphi\to \psi_0$ for $t\to\infty$.

\subsection{Two-parametric EOS}
\label{two-parametric_eos}
To illustrate how equation (\ref{eq16}) works in case of two-parametric EOS, we consider  $h(e,\varphi)=\xi e-U(\varphi)$, $\xi=const$,  which is obtained as a  generalization of (\ref{eq21}) by changing $\eta \to U(\varphi)$. We assume in this subsection $0<\xi<2$.

After substitution $\tilde V=V-U/2$ equation (\ref{eq16}) yields
\begin{equation}\label{eq_for_V-2-par}
\frac{d\tilde V}{d\varphi}=-S \sqrt{ 24\pi \tilde V\left[ (\xi\tilde V-\frac{2-\xi}{2}U(\varphi)   \right]},
\end{equation}
where  $S=\mbox{sign}(\dot\varphi)$.

This equation can be used either to derive $V(\varphi)$ for given $U(\varphi)$ or, vice versa, to find  EOS on condition that $V(\varphi)$  is given:
\[
U(\varphi)=\frac{2}{2-\xi}\left[\xi\tilde V-\frac{1}{24\pi\tilde V } \left( \frac{d\tilde V}{d\varphi}\right)^2   \right]
\] 
By considering various $\tilde V$ one can generate examples with subsequent verification of equation (\ref{eq_for_V-2-par}) and inequalities $\Theta_0>0,\,E_0>0$. We  give two such examples dealing with simple elementary functions.

(i) For  $\tilde V(\varphi)=A^2\varphi^2$, $A=const>0$, we have
\[
U(\varphi)=\frac{2 A^2}{2-\xi}\left(\xi \varphi^2-\frac{1}{6\pi}\right),\,
 V(\varphi)=\frac{2A^2}{2-\xi}\left(\varphi^2-\frac{1}{12\pi}\right).\]
Equation (\ref{eq_for_V-2-par}) is valid if $\dot \varphi(t)\varphi(t)<0$. The non-trivial particular solution satisfying both the equations of  H and SF models exists for $t<t_1=const$ 
\[ 
\varphi(t)=\pm \frac{A (t_1-t)}{\sqrt{3\pi(2-\xi)}},\quad e(t)=\frac{2A^4(t-t_1)^2}{3\pi(2-\xi)^2}.  
\]

(ii) The choice 
 $\tilde V(\varphi)=A^2\exp{(\alpha\varphi)}$, where $\alpha,A>0$ are constants, generates
\begin{equation*}
 \quad U(\varphi)=\frac{2A^2e^{\alpha\varphi}}{2-\xi} \left(\xi-\frac{\alpha^2}{24\pi}\right), \, V(\varphi)=\frac{A^2e^{\alpha\varphi}}{2-\xi}\left(2-\frac{\alpha^2}{24\pi}\right).
\end{equation*}
Equation (\ref{eq_for_V-2-par}) is valid if $\alpha\dot \varphi(t)<0$.
The particular solution of (\ref{eq8SF}) is 
\[
\varphi(t)=\frac{2}{\alpha}\ln\left[\frac{{4\sqrt{3\pi(2-\xi)}}}{A\alpha^2(t-t_1)}
\right], \,t>t_1=const.
\]
The corresponding solution of hydrodynamical equation (\ref{fluid}) is
$e(t)=e_f(t)=96\pi \alpha^{-4}(t-t_1)^{-2}$.

\section{Discussion}
It is clear that the hydrodynamic description of  DE is an oversimplification of the real cosmological situation in comparison with field-theoretic models. A consistent description of hydrodynamic phenomena assumes the local thermodynamical equilibrium. It is unclear how this assumption works as regards DE in the early Universe and in the modern era. Nevertheless, this does not prevent us from using the hydrodynamical model on a formal level by equating the scalar field energy momentum tensor to the hydrodynamical one. On the other hand, some solutions of hydrodynamic models that are widely used in cosmological considerations, can be interpreted in therms of the SF-models. 

In this paper we found conditions for the SF-model that make this possible in case of the homogeneous isotropic spatially flat cosmology, under additional restriction on the form of the SF Lagrangian to be a canonical one. This is a very restrictive  requirement; it leads lead to the differential equation for the potential $V(\varphi)$, which is effective on intervals with the constant $S=\mbox{sign}(\dot \varphi(t))$.  Moreover, the space of solutions of the SF-model is much wider than that of the barotropic H-model. In any case, the  global equivalence between H and SF models for all modes of cosmological evolution is impossible.  This is clearly seen in the examples of Section \ref{examples}. 

This, however, does not prohibit using the H-SF analogy to study some special regimes. The hydrodynamical solutions with EOS (\ref{eq21}) yield the SF solutions for the potential (\ref{potential_1}),  when SF rolls down the  potential well  or descents down the potential hill (Section \ref{examples}). But the H-model cannot describe the SF oscillations near minimum of the potential, though this regime being important for particle creation at the post-inflationary stage of the cosmological evolution \cite{Constraints_inflation}. On the other hand, some singular solutions  like the ``Big Rip'' \cite{caldwell_doomsday} that may take place for certain EOSs  are ruled out in case of the SF counterparts.

The restriction on the initial data reduces possibilities to use the hydrodynamical representation of the "restricted" SF model. This  trouble is mitigated by the possibility to investigate close solutions. We derived  conditions that ensure certain closeness  of the SF-model energy-momentum tensors to that of the H model. In this sense the fiducial solution, which satisfies equations of both H and SF models, well approximates nearby solutions and describes their qualitative properties.

\acknowledgments
This work has been supported in part by the Department of target training of Taras Shevchenko National University of Kyiv under National Academy of Sciences of Ukraine (project 6$\Phi$).

\appendix
\section{SF-H correspondence without restrictions}\label{general_case}
The hydrodynamical and scalar field approaches are equivalent,  if  
\begin{equation}\label{equal_Tmunu}
T^{(h)}_{\mu\nu}=T^{(sf)}_{\mu\nu} ;
\end{equation}
This equivalence can be used to find special solutions of hydrodynamics equations by means of the SF equations \cite{borshch-zhdanov}. However, 
(\ref{equal_Tmunu}) presupposes that the perfect fluid flow involved is a relativistic
analog of the classical potential flow \cite{borshch-zhdanov, diez}. Indeed, besides (\ref{h1}),  equations (\ref{equal_Tmunu}) yield 
\begin{equation}
\label{h1u_app} u_\mu F =\varphi _{,\mu }, \quad X>0,\quad F=\sqrt{\frac{e+p}{\partial L/\partial X}},
\end{equation}
where  $X>0$ is  a solution of equation (\ref{h2}) for given EOS. This can be easily shown by considering (\ref{equal_Tmunu}) in a locally Lorentz frame (where at some point $x_0$ we have $g_{\mu\nu}(x_0)= \eta_{\mu\nu}$, $\partial_{\alpha}g_{\mu\nu}(x_0)=0$), which is also an instantaneous proper frame for $u^{\mu}(x_0)=(1,0,0,0)$. 
 The inequality $X>0$ must be fulfilled because $u^{\mu}$ is timelike; therefore, the H-model cannot be equivalent to the SF model in case of a stationary SF.
On account of (\ref{sf-e-m-tensor},\ref{h-e-m-tensor}) and   \ref{h1u_app})   we get
\begin{equation}
\label{equiv_cond_0_app}
\partial_{\nu}[F(e,\varphi)u_{\mu }]=\partial_{\mu}[F(e,\varphi)u_{\nu }].  
\end{equation}

Usually for a given  EOS  $e=E(p,\varphi)$, the differential equation (\ref{h2})   has a solution $L$ that transforms (\ref{h2}) into an identity. On the other hand, for given $L(X,\varphi)$,  equations (\ref{h1}) represent the EOS parametrically, the domain of $E$ as a function of $p$ depending upon the range of $L$. In this sense we can speak about some equivalence of H and SF models, provided that conditions (\ref{equiv_cond_0_app}) are fulfilled; this latter depends on the initial conditions of the hydrodynamical problem. Obviously, these conditions may be not satisfied if we deal with an arbitrary hydrodynamic flow, i.e. there is no full ``equivalence'' between H and SF models.  However, the conditions (\ref{equiv_cond_0_app}) are always fulfilled in case of a  homogeneous cosmology, where the gradient $\partial_{ \mu} \varphi \ne 0$ is time-like and all the functions involved depend on the time variable only. In a more general case, the relativistic ideal fluid flows satisfying (\ref{equiv_cond_0_app}) may be considered as an analogue  of  classical irrotational flows.


\end{document}